\documentclass[]{article}
\usepackage{epsfig}

\newcommand{\be}{\begin{equation}}
\newcommand{\ee}{\end{equation}}
\newcommand{\bea}{\begin{eqnarray}}
\newcommand{\eea}{\end{eqnarray}}
\newcommand{\no}{\nonumber}

\begin{document}
\title{ {\bf Solar Neutrino Zenith Angle Distribution \\
         and Uncertainty in Earth Matter Density } }
\author{
Lian-You Shan and Xin-Min Zhang\\
Institute of High Energy Physics, \\
Chinese Academy of Science, P.O.Box 918, Beijing 100039, P.R. China   \\
}
\maketitle
\begin{abstract}
We estimate in this paper the errors in the zenith angle distribution for
the charged current events of the solar neutrinos caused by 
the uncertainty of the earth electron density.
In the model of PREM with a $5\%$ uncertainty
in the earth electron density 
we numerically calculate the corrections to the correlation between 
$ [N]_5 /[N]_2 $ and $ [N]_2 /[N]_3 $, 
and find the errors notable. 
\end{abstract}

Forthcoming results from SNO\cite{sno} include a measurement of the day-night
asymmetry
($A_{DN}$)\cite{dnask,regen,lisino,kerry}.
 This measurement
is crucial to confirming the
matter conversion solution
to the solar neutrino problem. 
And the analysis on
the zenith angle distribution of
the events during the night
may provide some insights to
distinguish the 
various MSW solutions, {\it i.e.} LMA, LOW and SMA\cite{gg}.
In the calculation of the regenerated
$\nu_e $ flux,
the electron density of the earth matter with which the neutrinos interact
is a critical quantity.  
The uncertainty in Earth-matter density and chemical component can be 
a major cause of the error in $A_{DN}$ and the zenith angle distribution.
So it will be interesting to estimate these errors. Furthermore
since the experimental value of the $A_{DN}$ is around
$\sim 0.047$\cite{dnask} and the theoretical expectations
on the zenith angle
distributions are small in
magnitude\cite{gg}, it is necessary to
perform a  
quantitative estimation on these errors.

In this paper, we follow the procedure outlined in [7] and study
the uncertainty in the earth matter 
density, then investigate its implications on the predictions of $A_{DN}$
and the zenith angle distributions. 
We quantify the uncertainties of the earth matter in terms of two
parameters: one is $\delta N_e / N_e $, the variation in magnitude of the
density which generally is expected to be around
a few percent; the second one is
$\delta x$ which specifies the limitation on the
spacial dimension
by geophysics experiments and inverting calculations used in the fit of
the earth density models.
In general the scale $\delta x$ is not much larger than 
the neutrino oscillation length, e.g., in the case with the
parameters of the favored LMA
solution, so its 
effect might arise beyond the linear
order. We will show in this paper that this effect causes an 
 sizable error 
in the zenith angle distributions.

To begin with 
we consider a two-neutrino mixing model for simplicity. As discussed
in\cite{liuqy,lisino}
the neutrino can be treated as a incoherent mixture of two 
mass eigenstates. 
In the day-time the survival probability for $\nu_e$ is given 
by,
\be
P^D = P_1 \cos^2\theta + ( 1 - P_1 ) \sin^2\theta ~, 
\ee
where the mixing angle is defined through,
\be
\nu_1 = \cos\theta ~\nu_e - \sin\theta ~\nu_\mu ~,~
\nu_2 = \sin\theta ~\nu_e + \cos\theta ~\nu_\mu ~, ~
\ee 
and $P_1$ is the probability of the $\nu_e \to \nu_1 $ conversion 
inside the sun\cite{petc,gg}. 
During the night time, the presence of the earth matter leads to   
a zenith angle dependent regeneration of the $\nu_e$,
\be
P^N = P_1 + ( 1 - 2 P_1 ) P_{2e} 
 = P^D - 2 X f_{reg} ~,
\label{night}
\ee
where $P_{2e}$ is the probability of
the $\nu_2 \to \nu_e $ conversion inside the Earth,  
$ X = P_1 - 1/2 $. And  
\be
f_{reg} (\theta_z) \equiv P_{2e} ( earth~matter ) - P_{2e} ( vacuum ) ~,   
\ee
is the regeneration factor
which vanishes in the absence of the Earth matter effect.
Defining ${\bar f}_{reg}$ as the regeneration factor 
integrated over the zenith angle, one has the day-night asymmetry,
\be
A_{DN} \equiv  \frac{ P^N - P^D }{ \frac{1}{2} ( P^N + P^D ) }
= \frac{ - 2 X {\bar f}_{reg} }{ 0.5 + 
( \cos 2\theta - {\bar f}_{reg} ) X }.
\label{dna}
\ee
The matter effects has entered the day-night asymmetry through $f_{reg}$. 
Formally,
\be
P_{2e} ( E_{\nu} , \theta_z ) = { {\Big |}
 \cos\theta \lfloor {\cal T} exp \lfloor -i \int_0^{D\cos\theta_z} 
H[ N^{\theta_z}_e(x) ] dx \rfloor \rfloor_{ee} + 
\sin\theta \lfloor {\cal T} exp \lfloor -i \int_0^{D\cos\theta_z}
H[ N^{\theta_z}_e(x) ] dx \rfloor \rfloor_{e\mu} 
{\Big |}  }^2 ~,
\ee
where $D=12742$ is the diameter of the Earth in unit of kilometer and 
$H [N^{\theta_z}_e(x) ] $ is the effective Hamiltonian for the
given trajectory 
with zenith angle $\theta_z $, 
\be
H[N^{\theta_z}_e(x) ] = \frac{ \Delta m^2}{4 E_\nu}
\left( \begin{array}{cc}
2 \sin^2\theta & \sin 2\theta \\
\sin 2\theta & 2 \cos^2\theta 
\end{array}   \right)  +
\left( \begin{array}{ccc}
\sqrt{2} G_F N^{\theta_z}_e(x) & 0 \\
0 & 0 
\end{array} \right) .
\label{heisenberg}
\ee
In Eq.(\ref{heisenberg}), 
$N^{\theta_z}_e(x) $  is the Earth Electron Density (EeD) along the
trajectory of the
zenith angle $\theta_z$. If the density is known, the
regeneration factor $ f_{reg} = P_{2e}- \sin^2\theta$ 
can be calculated accurately. As an example we take the Preliminary
Reference Earth Model (PREM)\cite{prem} and plot in Fig.1.(a)
the regeneration factor 
as a function of the zenith angle. In the numerical calculation we take
the neutrino energy to be 11 MeV and
the oscillation parameters to be\cite{glob}
\bea
LMA ~: ~ \Delta m_{12} = 3.7 \times {10}^{-5} , \tan^2\theta = 0.37 ~, \no
\\
LOW ~: ~ \Delta m_{12} = 1.0 \times {10}^{-7} , \tan^2\theta = 0.67
\label{snopara}
\eea
One can see from this figure that the regeneration factors oscillate
periodically with certain lengths. And different oscillation lengths
correspond to different MSW solutions. 

Given the parameters in Eq.(\ref{snopara}) and the standard solar
density\cite{soden}, we follow \cite{gg,petc} and obtain numerically that 
$\cos 2\theta_S\approx -1 $
and $P_c \approx 0 $, which can be used to get the $\nu_e\to\nu_1$
conversion in the Sun.
Fluctuations in the solar density will affect $P_1$, consequently
also influence the MSW solutions\cite{solaflu}. 
In that situation a variance of $P^D$ has been defined to estimate 
the relevant error \cite{solarvar}.
In this paper, however, we concentrate on the errors caused 
by the uncertainty of EeD.

The EeD available today, is known only to some certain precision 
\cite{ohson,hara}.
As to PREM, significant uncertainties due to the local variation
have been documented\cite{cases}. 
Quantitatively its precision is roughly
$5\%$ averaged per spherical shell with
thickness of 100 Km or so \cite{bolt}.
 The uncertainties of the Earth matter density
cause 
errors in the calculation of the $\nu_e$ survival probability during
the night time. In the following we study numerically 
the uncertainties in the solar
neutrino zenith angle distributions.

As described detailly in \cite{denxx}, we introduce a weighted average
over the whole sample space of possible earth density profile. Denoting
the averaged earth density function, such as the widely used PREM 
by $\hat{N_e}(x)$, 
we have
   $\hat N_e(x) = < N_e(x) > = \int [DN_e]F[N_e(x),x] [{\cal D} N_e]$
where
$F[N_e(x),x] [{\cal D} N_e] $ is the probability of obtaining the EeD
$N_e(x)$ in the neighborhood of x, 
\bea
& &F[ N_e(x) , x ] = \frac{1}{ N_e (x) \sqrt{2 \pi} s( x ) }
exp\{ -  \ln^2 \lfloor ~N_e(x) / N_0(x) ~\rfloor
 / \lceil 2 s^2 ( x ) \rceil 
\}, \no  \\
& & s(x) = \sqrt{ \ln \lceil 1 + r^2(x) ~\rceil },~~
N_0(x) =  {\hat N}_e(x) exp \lfloor -s^2(x)/2 ~~\rfloor ,
\label{eqlogaus}
\eea
where
$r(x) = \sigma (x) /{\hat N}_e(x) $
characterizes the precision of the earth electron density.

The averaged value and the variance of 
the $\nu_2 \to \nu_e $ conversion probability 
can be written now separately as,
\bea
< P_{2e}(E_{\nu} , \theta_z) > & \equiv & 
\int P_{2e}(\theta_z) F[ N^{\theta_z}_e(x) ] 
                 [ {\cal D}  N^{\theta_z}_e  ] ~ \no  \\
&=& \lim_{I\to\infty} \int
\prod_{i=1}^I F[ N^{\theta_z}_e(x_i),x_i ]
d N^{\theta_z}_e( x_i ) \times    \no \\
&~&~~~~~P_{2e}(\theta_z)
[ \{ N^{\theta_z}_e( x_1 ), ... N^{\theta_z}_e( x_i ), ... 
N^{\theta_z}_e( x_I) \} ]  \no  \\
& = & \lim_{K\to\infty} K^{-1} \sum_{k=1}^K
{ \tilde P_k } ( N^{(k)}_e ) ,   \no  \\
\delta f_{reg} &=&  \delta P_{2e}(E_{\nu} , \theta_z) \equiv 
\sqrt{ < P^2_{2e}(E_{\nu} , \theta_z) > - 
     { < P_{2e}(E_{\nu} , \theta_z) > }^2  }    \no   \\
&=& { [ \lim_{K\to\infty} {(K-1) }^{-1}
\sum_{k=1}^K { {\big ( } { \tilde P_k } - <  P_{2e} > 
{\big ) } }^2 ] }^{1/2}.
\label{vf}
\eea
We evaluate the functional integrations in Eq.(\ref{vf})  
using a method
similar to that of the lattice gauge theory. In the numerical
calculation we discretize the neutrino path 
into I bins, $\delta x_i$ (i = 1, 2, ...I) and in
the i-th bin the EeD function
$N_e(x)$ is given by Eq.(\ref{eqlogaus}). Furthermore we have replaced
the functional integration over the EeD by a sum over K arrays,
$N^{(k)}_e$ k=1, 2, ...K. In Eq.(\ref{vf})  
${ \tilde P}_k$ is the conversion probability evaluated with  
the $k'th$ density profile $N^{(k)}_e$ .
As to PREM 
each point in the array 
$N^{(k)}_e $ which consists of $ N_e(x_1), ....N_e(x_i), .... N_e(x_I)  $  
is generated from PREM weighted
with a Gaussian-like logarithm distribution.
Since the deviation from PREM due to local
variation is roughly $5\%$
and the deviation is averaged per spherical shell with thickness of
100 Km, we take $r = 5\%$ and choose the bin sizes $\delta x_i$ to be the
distance the neutrino travels along the path of zenith angle $\theta_z$
within a spherical shell of thickness 100 Km. So in general
 $\delta x_i$ will not be equal except
for $\theta_z = 0$.

We note that the EeD uncertainty scale $\delta x$ differs
from the one, $l_\rho = \rho /{d \rho \over dx}$ considered in 
\cite{gg} to
characterize the flatness (adiabaticity) of the density profile. Both of
these scales are important to the studies on the neutrino oscillations in
matter. The effects of the $l_\rho$ can be taken into account in
the exact numerical calculation, however to reduce the error caused by $\delta
x$ a more precise density profile is needed. Especially when $\delta x$ is
comparable to the neutrino oscillation length in matter, 
one has to be careful in estimating the errors for oscillation probability.

In Fig.1.(b) we plot $\delta f_{reg}$ and $f_{reg}$ 
as a function of the zenith angle.
One can see from this figure that LMA suffers a larger error.
For LOW the error is roughly two percent and for SMA the error is
much smaller.
So we have not shown them in the figure.
Integrated over the zenith angle, 
it gives rise to a correction of $20\%$ roughly to the
$A_{DN}$ for LMA,
however the corrections
are small 
for LOW and SMA. 
Combined with Fig.1.(a), we see that the errors are small so
that  
$A_{DN}^{LMA}$ and $A_{DN}^{LOW}$ can be distinguished.

To see the effects on the solar neutrino observations,
we now estimate the errors in the rate of the charged current events 
during the night time.
Following \cite{gg} we define the normalized rate 
of the charged current events as,
\bea
[ CC ] (\theta_z) &\equiv& N_{CC} / N^{SSM}_{CC}   \no  \\
&=& \int_{E_{th}} dT_e 
\int_{E_{\nu} } d E_{\nu} P^N ( E_{\nu} , \theta_z ) \Phi ( E_{\nu} )
\int dT^{'} d \cos\theta_L 
{\hat {\sigma }} ( E_{\nu} , T^{'} ,  \cos\theta_L ) R ( T_e , T^{'} ) 
/ N^{SSM}_{CC}  \no  \\
&=& \int_{E^0_{\nu} } d E_{\nu} 
\Phi ( E_{\nu} ) P^N ( E_{\nu} , \theta_z ) 
\sigma_{CC} ( E_{\nu} ) /
N^{SSM}_{CC} ~~,  
\label{cc}
\eea
where $\Phi ( E_{\nu} )$ contains both the neutrino flux from the Boron
decay
and the $ He + P $ chain in the sun\cite{bahcall, kerry},  
and $ N^{SSM}_{CC}$ is the normalization factor 
which equals to the integral in $r.h.s.$ taken at $P^N=1$.
From the 2'nd to the 3'rd line of Eq.(\ref{cc}), 
the integration of the differential cross section 
${\hat {\sigma }}$ with respect to the recoil electron kinetics $T_e$ 
and the scatting angle $\theta_L$ has been replaced by 
a total charged current cross section ${\sigma }_{CC}$ of the neutrino on
the Deuteron, since the possible uncertainty from 
$T_e, ~\theta_L$ can be canceled in the $[CC]$ as a ratio
of $N_{CC}$ to $N_{CC}^{SSM}$. 
The $E_{\nu}$ dependence in  ${\sigma }_{CC} $ 
is accessed by employing a {\it {quick function}} from {\it {interpolation}} 
in \cite{kninterpo}.
The starting point of the neutrino energy is set at 
$E^0_\nu \approx Q + E_{th}$, with $Q= 1.442 MeV$ being 
the deuterium threshold energy 
and $E_{th} = 5 MeV$ the electron threshold energy.
In Fig.2, we plot the zenith angle distribution of the 
charged current events rate in Eq.(\ref{cc}).
One can see that
the 
SNO charged current data lies in the middle between the 
LMA and LOW. This serves also as a check of our numerical
calculation.
Following the binning method of \cite{gg},
we plot in Fig.3.(a) and Fig.3.(b) the charged current events
{\it v.s.} bins (note for the fifth bin $\cos\theta_z \sim ( 0.83,~
0.92 )$ in the case of SNO), which shows that,
\bea
LMA ~: ~ [N]_1 < [N]_2 \leq [N]_3 \leq [N]_4 \leq [N]_5 ,  \no \\
LOW ~: ~ [N]_2 \geq [N]_4 > [N]_1 \sim [N]_3 > [D]. 
\eea
Quantitatively, it reads,
$[N]_5 / [N]_2 =0.999 \approx 1$ and $ [N]_2 / [N]_3 = 0.995 \approx 1 $ 
for LMA while 
$[N]_5 / [N]_2 = 0.982 , [N]_2 / [N]_3 = 1.053 $ for LOW, from which 
it might be possible to distinguish
LMA from LOW. The calculation for SMA can be easily worked out, however
for simplicity we will not repeat it here.

Making use of Eqs.(\ref{cc}), (\ref{vf}), and (\ref{night}) we estimate the errors in the
charged
current events rate caused by the uncertainties in the earth electron
density
\be
\delta [CC] \propto \int_{E^0_{\nu} } d E_{\nu}
\Phi ( E_{\nu} ) ( -2 X \delta f_{reg} )  \sigma_{CC} ( E_{\nu} ) ,
\ee 
which we show in Fig.(\ref{niflu}) by the error-bars.
To avoid multi-fold integration which is a computer time consuming, 
we investigate $\delta f_{reg}$ at neutrino energies of $8, 10, 11, 12$
MeV and find the results almost unchanged.
To be conservative we have used the maximal value for $\delta f_{reg}$.

We see from the figure that the errors become larger as the zenith angle
increases in the case of LMA.
Averaged over bins we have
$ ( [N]_5 - \delta [N]_5 )/ [N]_2 \approx 0.935 $ while 
$( [N]_2 + \delta [N]_2) / ( [N]_3 - \delta [N]_3 ) \approx 1.043 $. 
As indicated in the figures 13-16 of \cite{gg} that
the LMA sheet in their correlation figures
mainly stretched along the $A_{DN}$ direction, 
we study a correlation between 
$[N]_5 /[N]_2 $ and $[N]_2 /[N]_3 $, which we show in Fig.5.
One sees that the point ($1,~1$) for LMA is swollen into a rectangle 
close to the point ( $0.982 , 1.053$ ) for LOW.
In this figure we have not shown the error bars for LOW since they are
small.

So far the precision of PREM which we assume is $5\%$. Certainly errors
on the zenith angle distribution become larger if the uncertainty in the 
earth electron density is bigger. Sure a
modern Earth's density model with higher precision 
will reduce the errors considered in this paper. As an example we take
density model AK135\cite{ak135}.     
The precision of AK135 is widely considered to be about $1\sim 2\%$, 
and its uncertainty scale is roughly $\delta x \approx 50$ Km 
since the model was presented in a data table.
Taking a $2\%$ uncertainty in the electron density we show our results in
Figure. 6. 
one finds 
$ ( [N]_5 - \delta [N]_5 ) / ( [N]_2 + \delta [N]_2 ) \approx 1.033 $ while
$( [N]_2 + \delta [N]_2) / ( [N]_3 - \delta [N]_3 ) \approx 1.017 $.
From Fig.6.(b), we see the gap between LMA and LOW enlarged. This
makes it easier to distinguish the LMA from LOW than the prediction from PREM.

In summary, we have estimated in this paper the errors in the zenith angle
distribution of the charged current event rates of the solar neutrinos
originated from the electron density uncertainty.
Our results show that the corrections are not significant in
the case of LOW and
SMA, 
however, error is notable for LMA.
Even though our estimations are given for specific parameters and
qualitatively, the results of this paper indicate that to observe the
zenith angle distribution 
a precise knowledge on the Earth electron density is necessary.

The work is supported in part by the NSF of China
under Grant No 19925523
and also supported by the Ministry of Science and Technology of China
under Grant No NKBRSF G19990754.

\begin{figure}[t]
\vspace{1.0cm}
\begin{center}
\epsfig{file=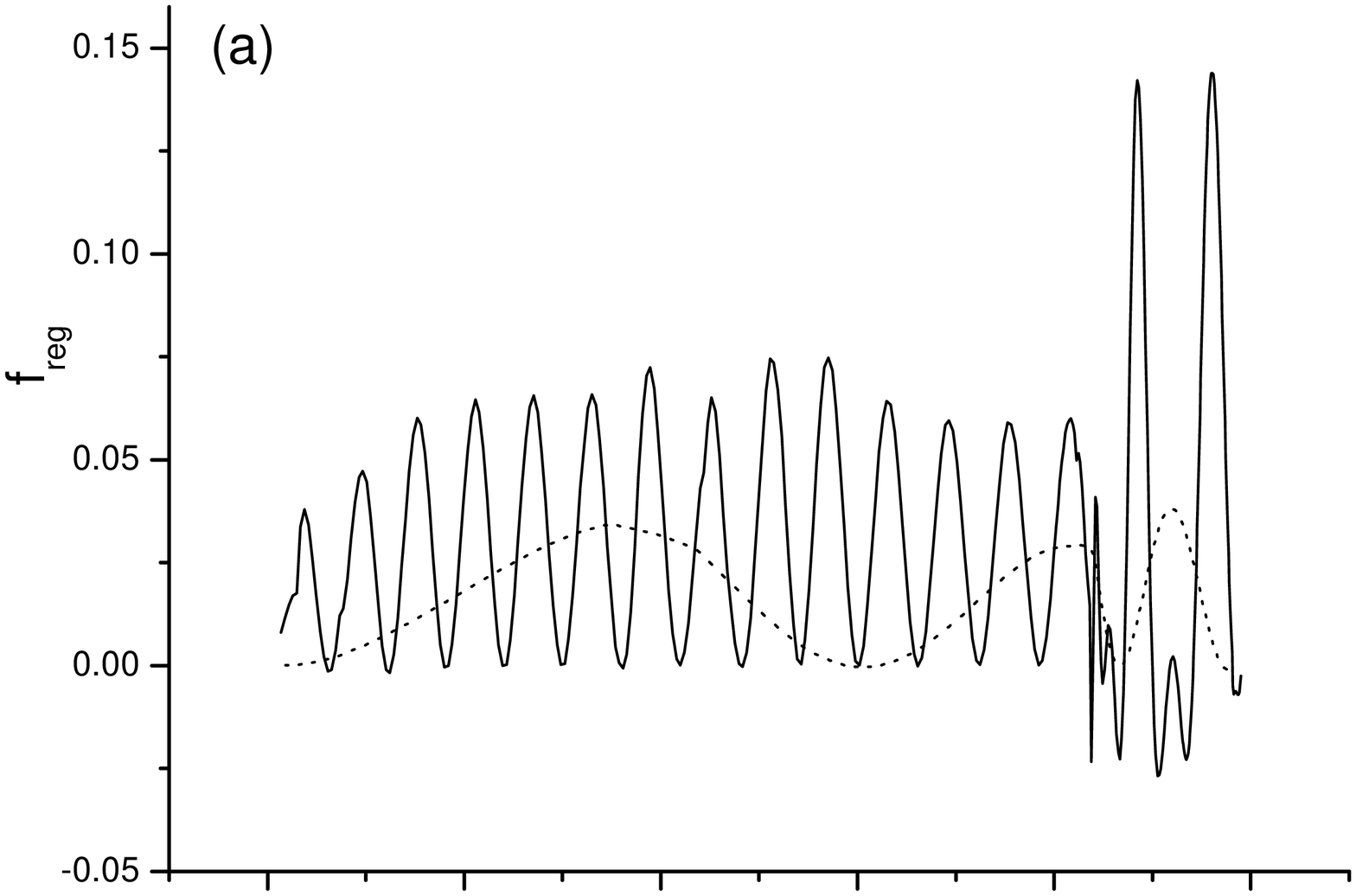,width=10cm}
\epsfig{file=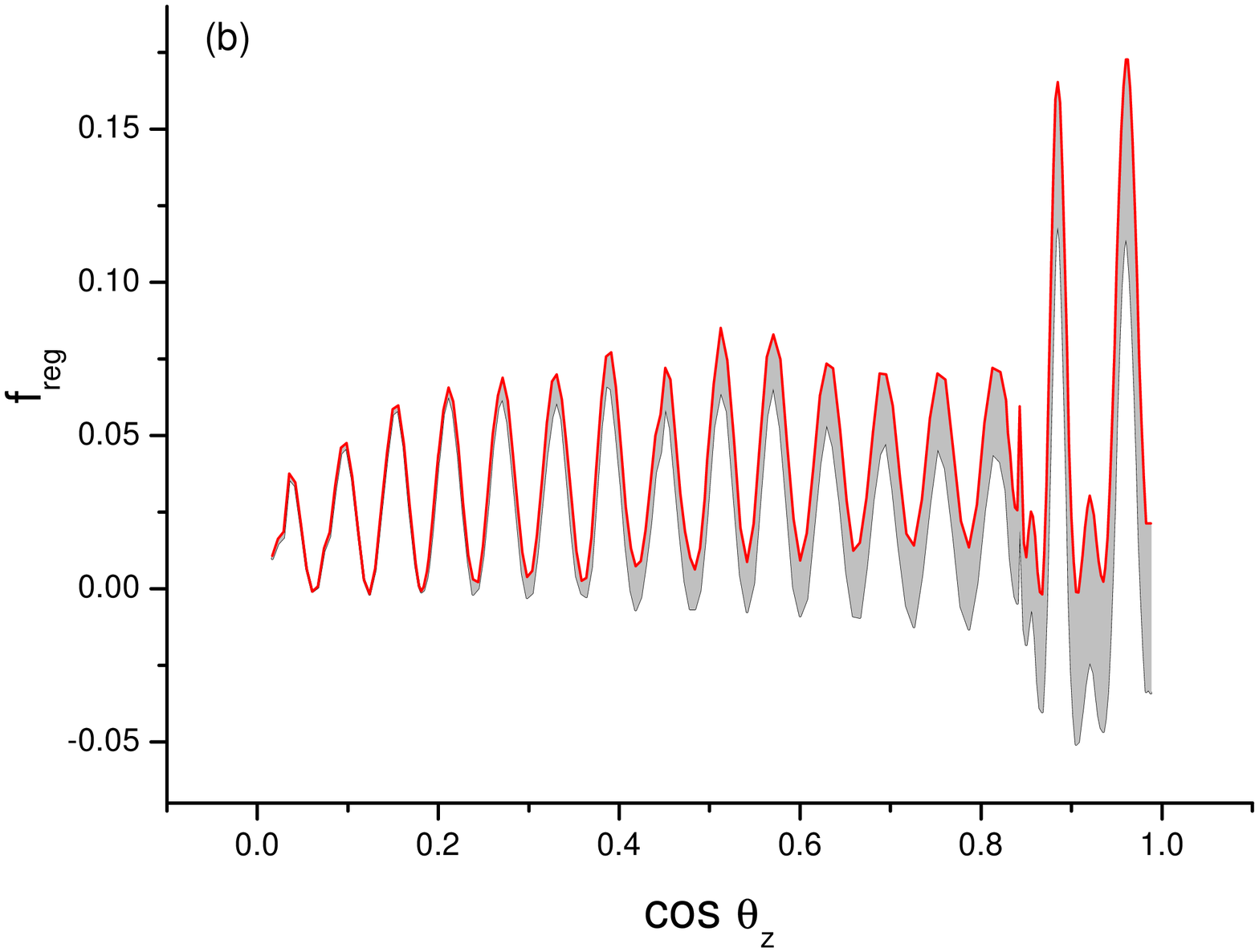,width=10cm}
\caption{Plot of the regeneration factor vs the zenith angles for 
neutrino energy at 11 MeV. The Earth matter model of PREM and the neutrino
oscillation
parameters in Eq.(\ref{snopara}) have been used.
(a) The solid line is for LMA while the dotted line for LOW.
(b) the error bars corresponds to the corrections due to the 
$5\%$ uncertainty 
in the matter density ( PREM ) .
The fluctuation in LOW case is smaller than the 
LMA case, so we have not shown them explicitly in this figure.
}
\label{freg}
\end{center}
\end{figure}

\newpage
\begin{figure}[t]
\begin{center}
\epsfig{file=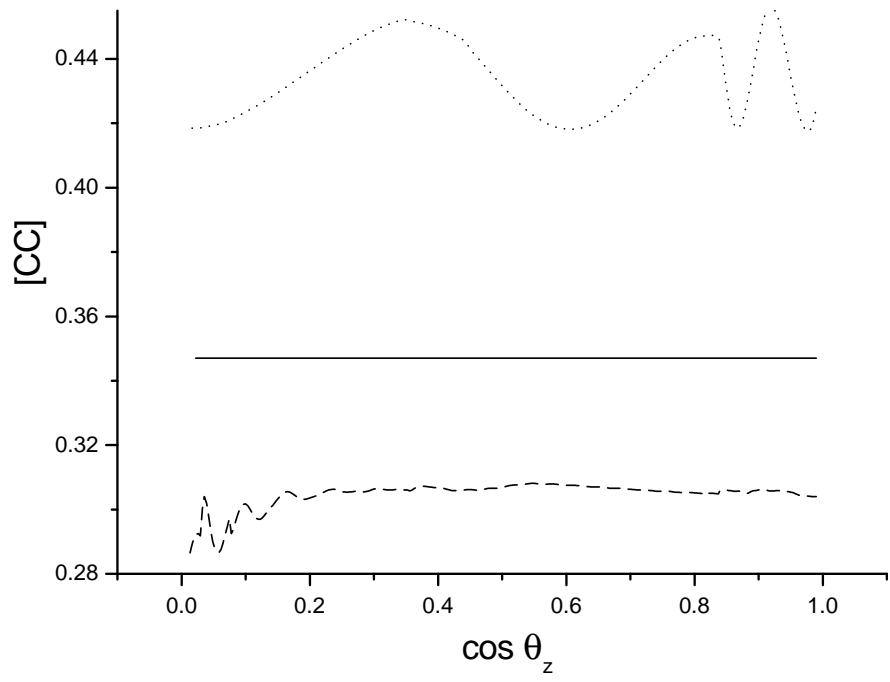,width=15cm}
\caption{ Charged current event rates vs the zenith angles.
The dotted line is for LOW and the dashed line for LMA.
The solid straight line is the data of the SNO observation.
}
\end{center}
\end{figure}

\begin{figure}[]
\begin{center}
\epsfig{file=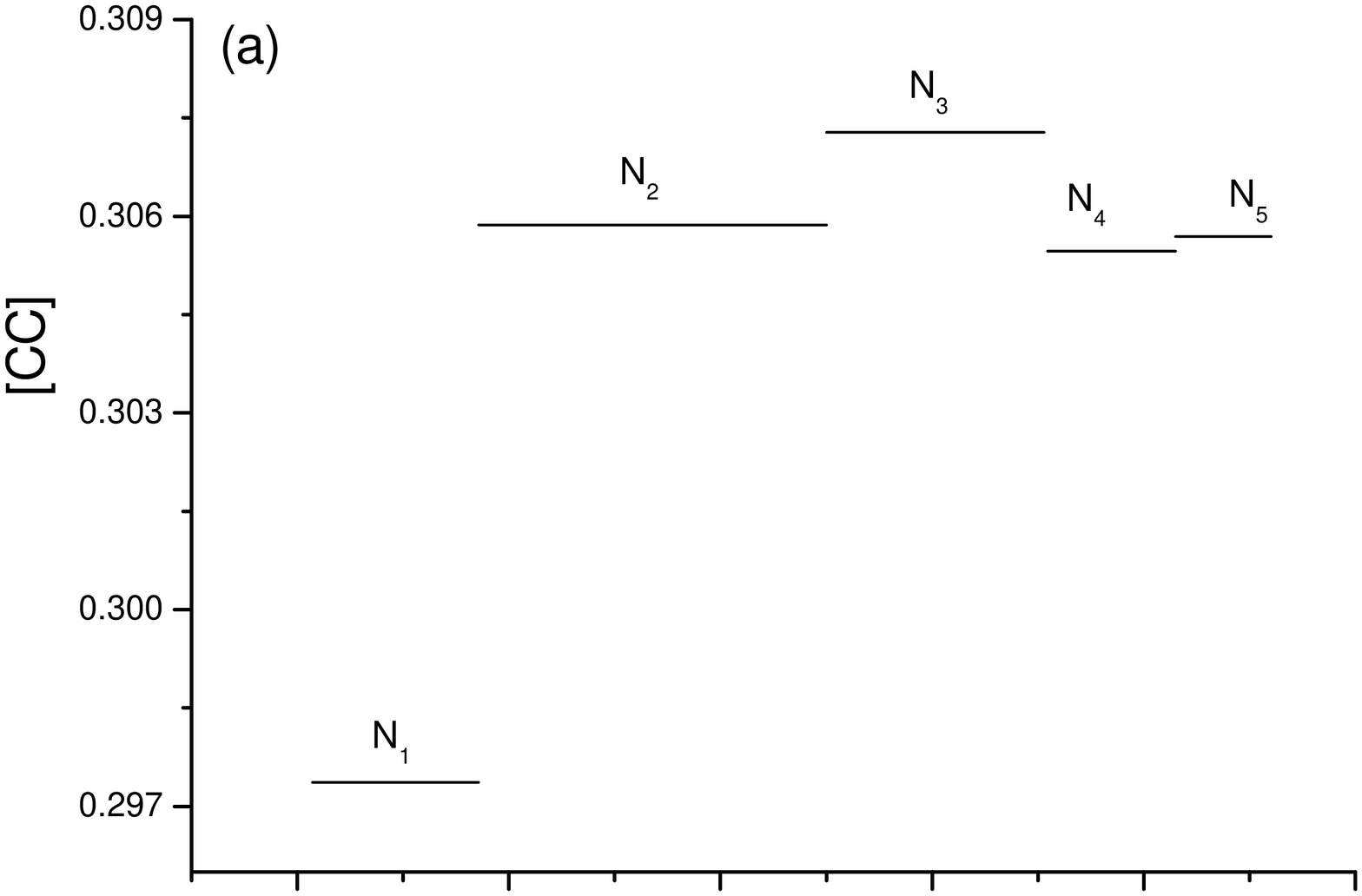,width=10cm}
\epsfig{file=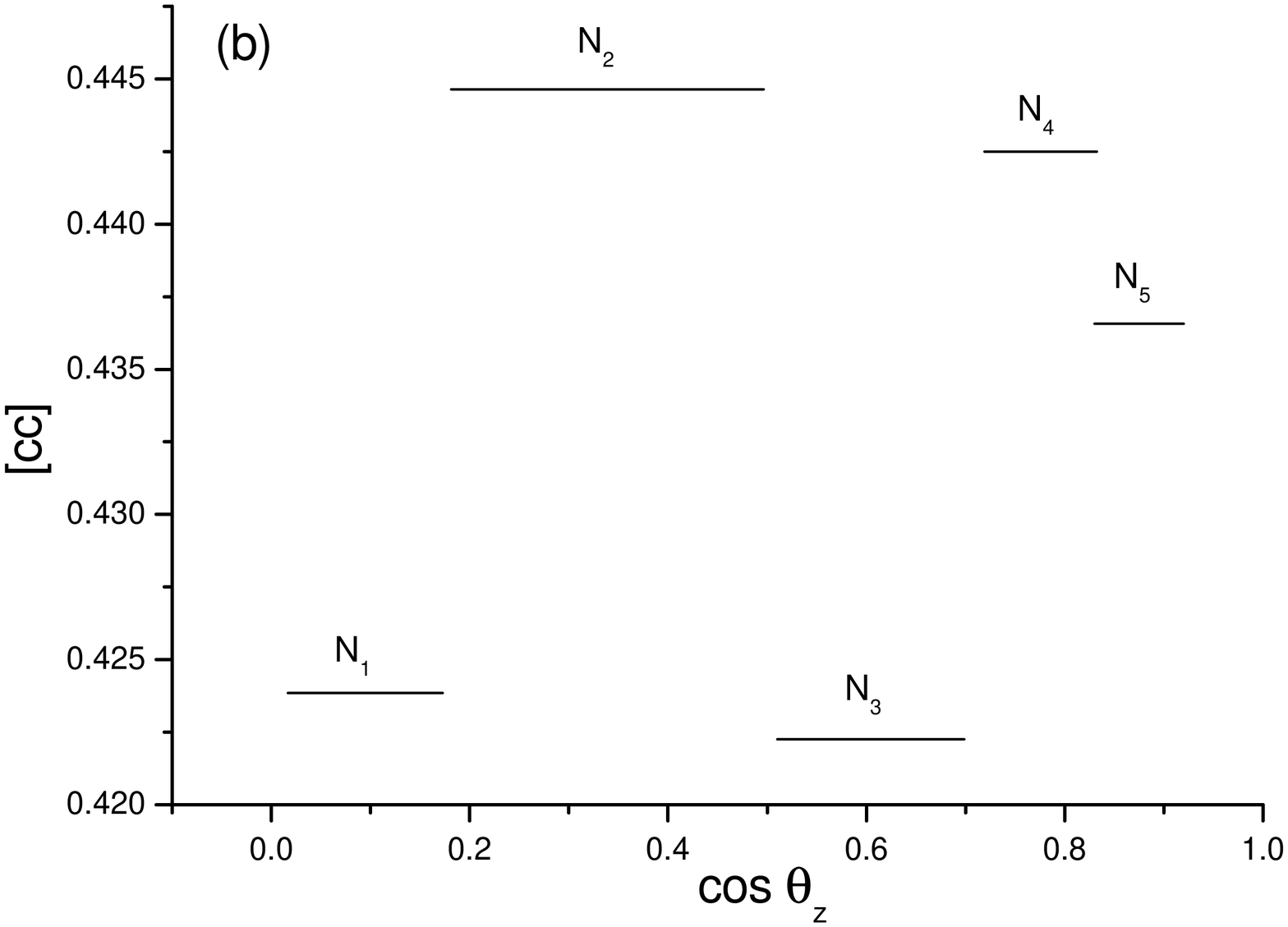,width=10cm}
\caption{ 
Plot of the charged current event rates averaged over bins as a function
of the
zenith angles.
(a) is for LMA  which corresponds to the dashed line in Fig.2, 
(b) is for LOW  corresponding to the dotted line in Fig.2. 
}
\label{smwnight}
\end{center}
\end{figure}

\newpage
\begin{figure}[t]
\begin{center}
\epsfig{file=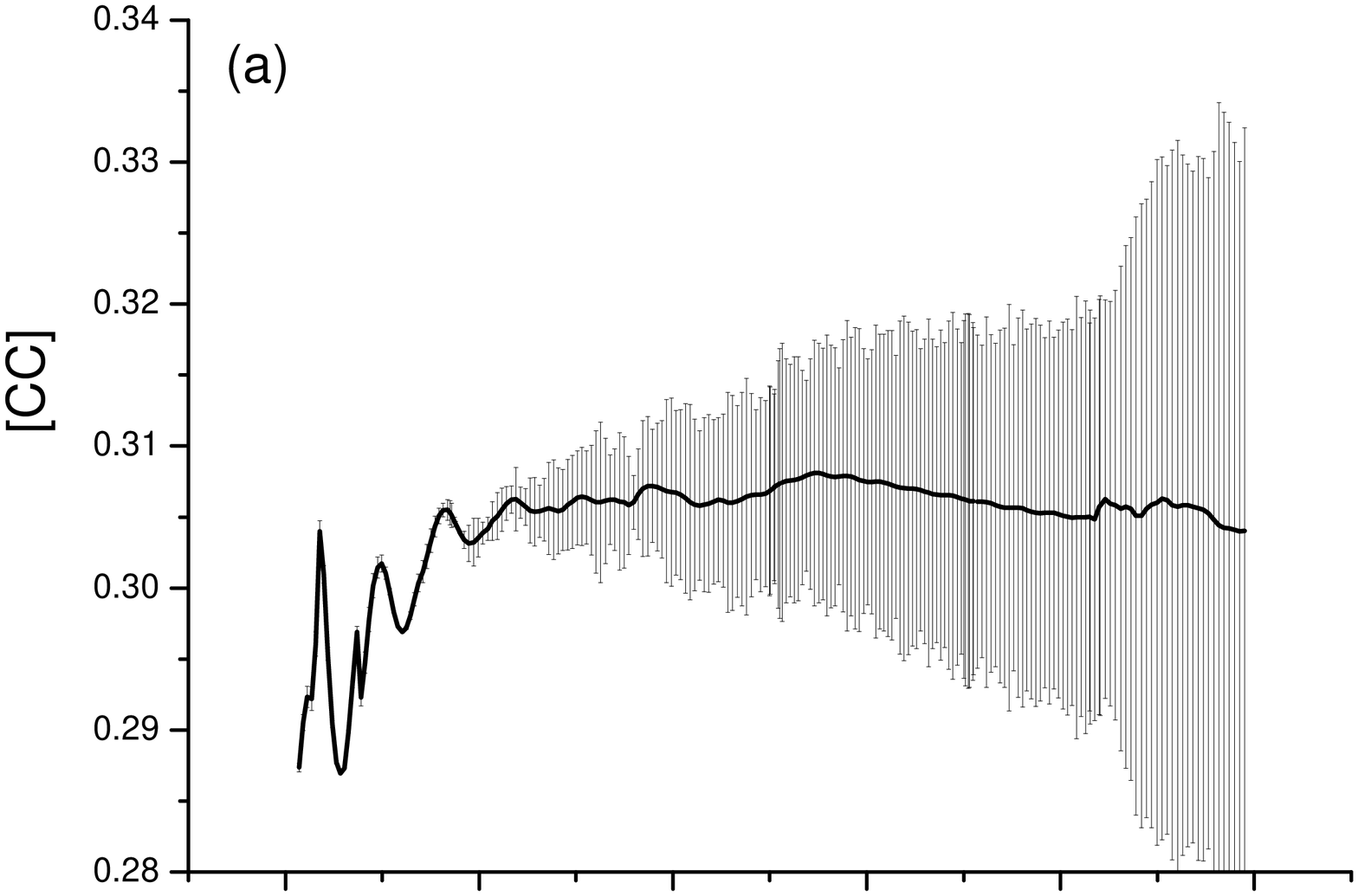,width=10cm}
\epsfig{file=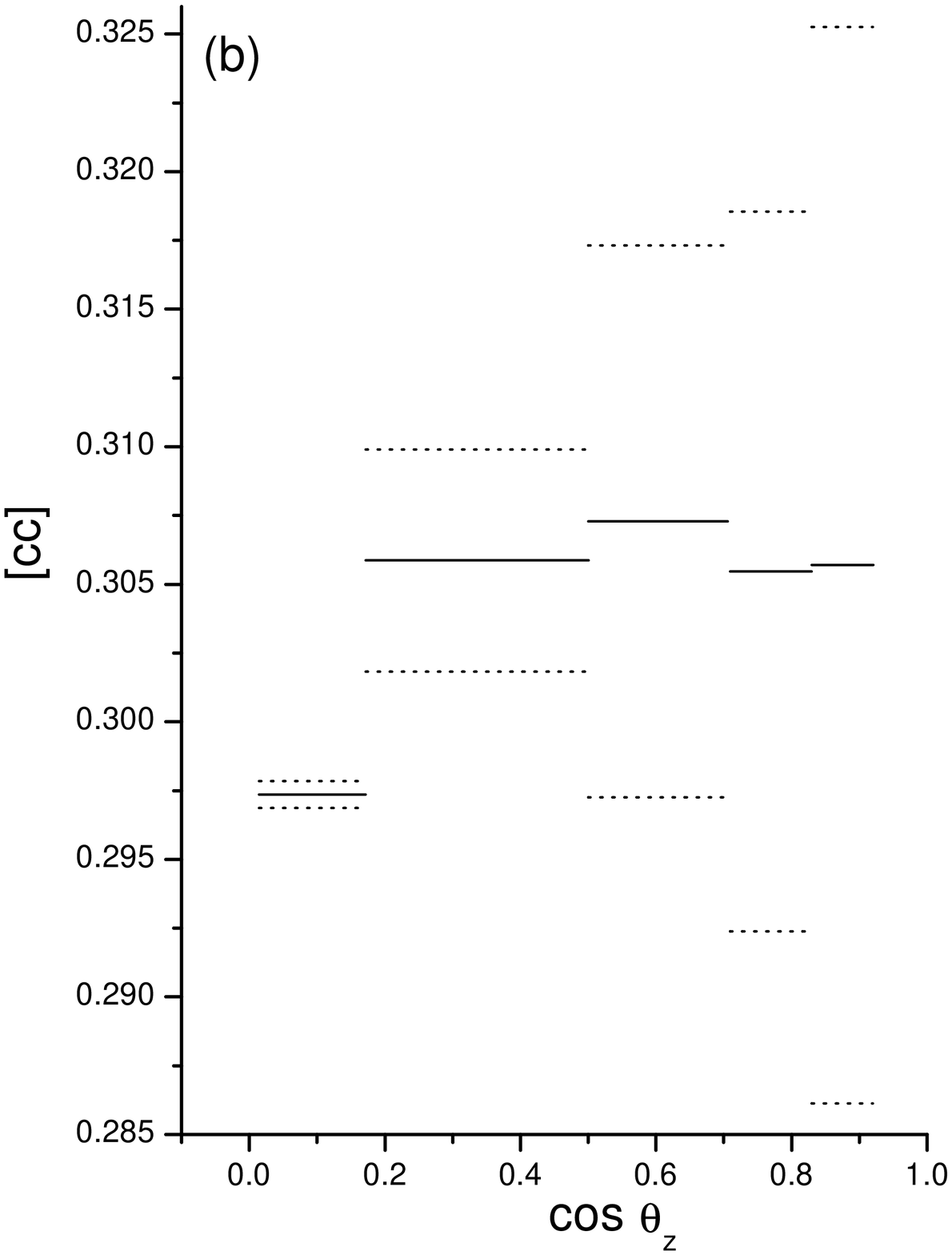,width=10cm}
\caption{ 
Plot of the
errors in the charged current event rates for LMA vs the zenith angles.
(a) shows the error-bars attached on the dashed line of Fig.2.
(b) The solid line is the same as that in Fig.3(a). And between the dotted
lines are the errors caused by the uncertainty in the electron density.
}
\label{niflu}
\end{center}
\end{figure}

\newpage
\begin{figure}[t]
\begin{center}
\epsfig{file=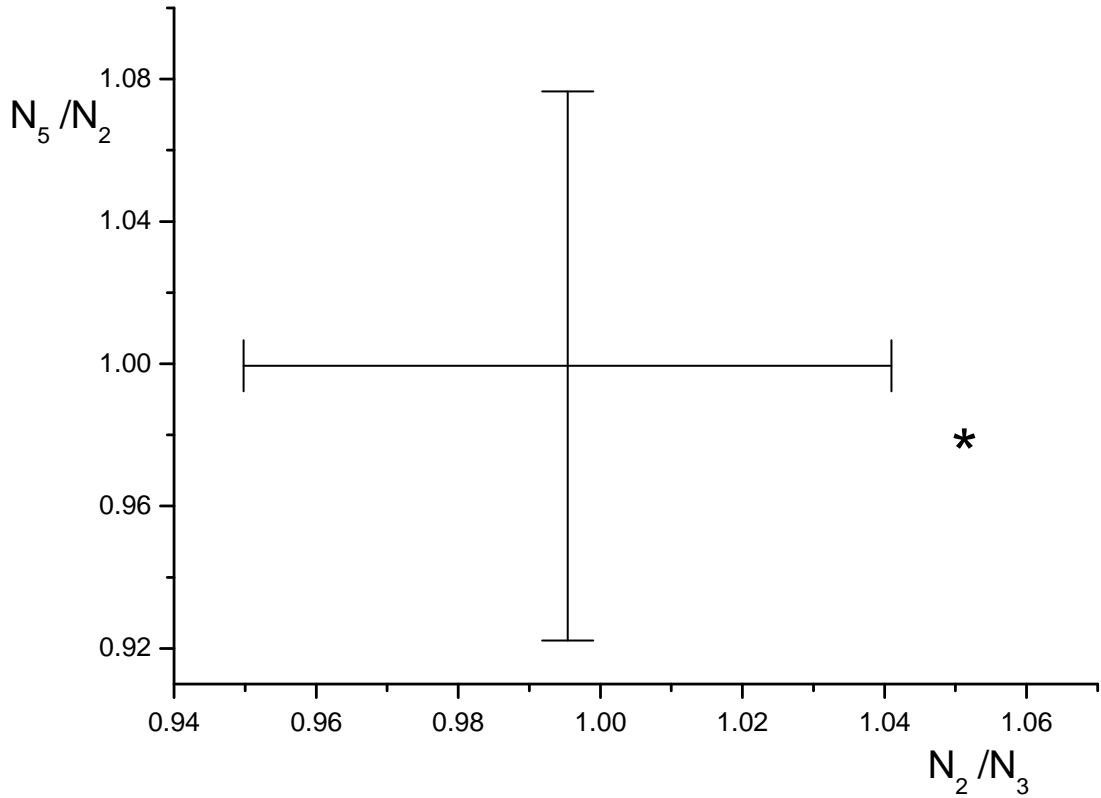,width=18cm}
\caption{ Plot of  
the correlation between  
$N_5 /N_2$ and  $N_2 /N_3$ . 
The center of the cross corresponds to the best-fit LMA, 
the star is for the best-fit LOW . 
The error bars ( cross ) span a rectangle and indicate a possible 
blur due to the uncertainty of EeD.
}
\label{correla}
\end{center}
\end{figure}
 
\newpage
\begin{figure}[t]
\vspace{1.0cm}
\begin{center}
\epsfig{file=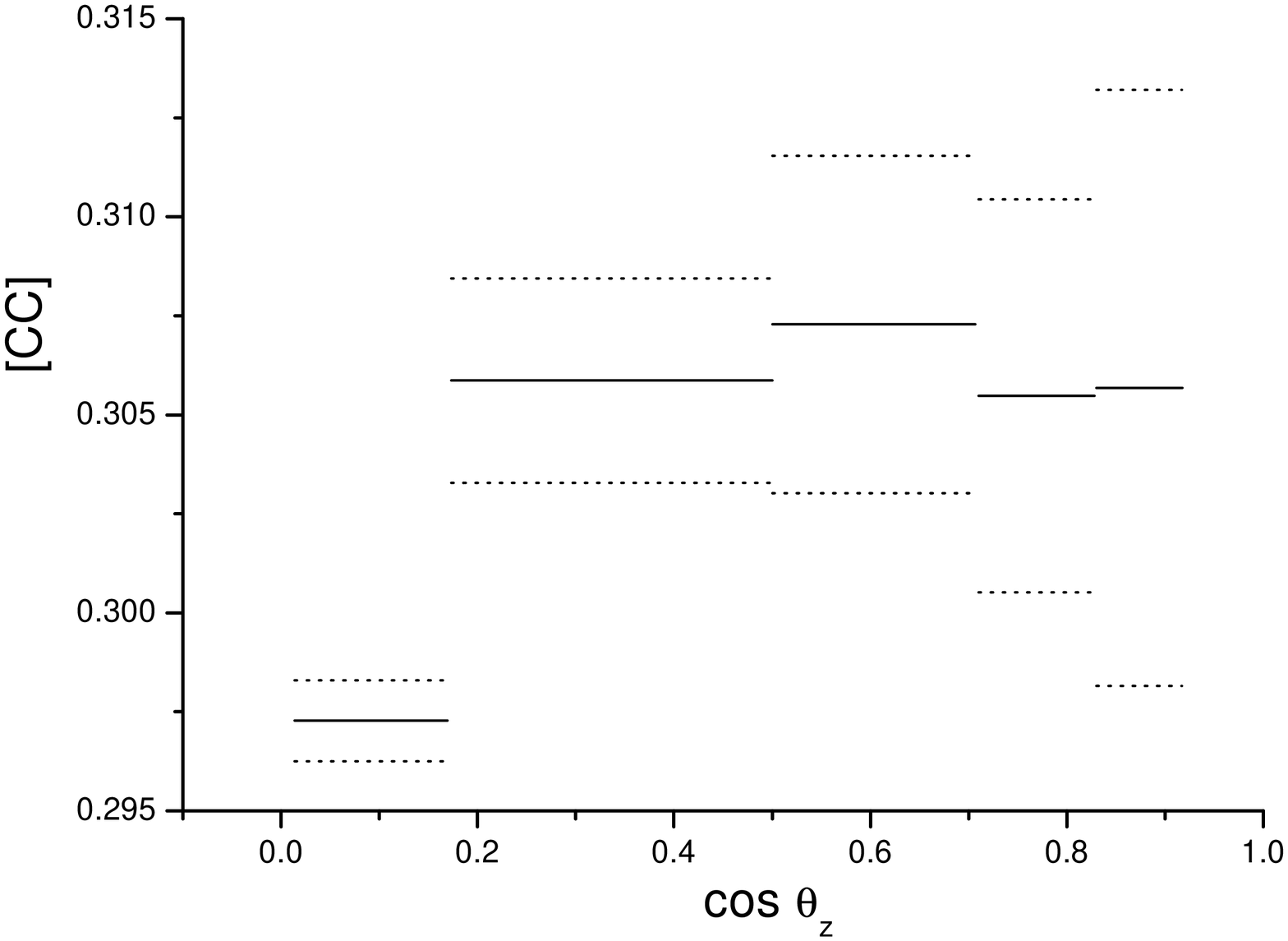,width=10cm}
\epsfig{file=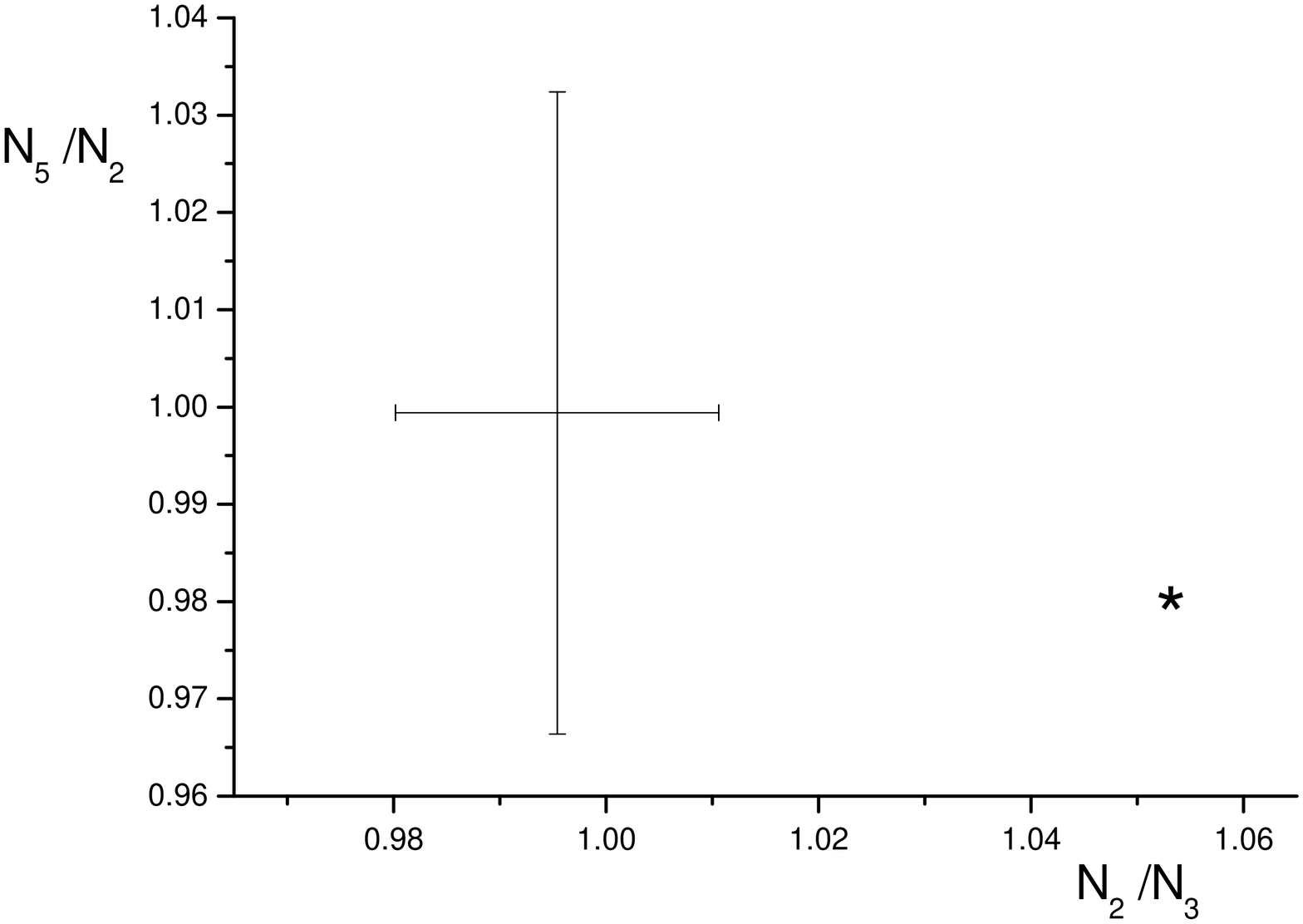,width=10cm}
\caption{(a) is the same as Fig.4.(b), but with a $2\%$
EeD uncertainty in the AK135 model.
(b) the same as Fig.5. but with AK135 instead of PREM.
}
\label{akp2}
\end{center}
\end{figure}

\end{document}